\documentstyle[preprint,psfig,aps]{revtex}
\newcommand{\ignore}[1]{} 
\begin{document}
\draft
\title{Pairing in low-density Fermi gases}
\author{T. Papenbrock and G.F. Bertsch}
\address{Institute for Nuclear Theory\\
University of Washington, Seattle, WA 98195 USA
}
\maketitle
\def\eq#1{eq. (\ref{#1})}
\def\be{\begin{equation}}
\def\ee{\end{equation}} 
\def\ba{\begin{eqnarray}}
\def\ea{\end{eqnarray}} 
\begin{abstract}
We consider pairing in a dilute system of Fermions
with a short-range interaction. While the theory is ill-defined for
a contact interaction, the BCS equations can be solved in the
leading order of low--energy effective field theory.  The integrals
are evaluated with the dimensional regularization technique, giving
analytic formulas relating the pairing gap, the density, and the 
energy density to the two-particle scattering length.
\end{abstract}

In the theory of fermionic matter, the expansion about the low--density
limit has been invaluable for understanding the structure of the
theory and the role of the interaction.  At low densities, the
interaction needs only be characterized by its scattering length
to get expansions for the energy density, excitation spectrum,
etc.\cite{ab63}.  However, to our knowledge the pairing singularity has
never been incorporated into this framework.  We have for
example only the qualitative statement in ref. \cite{ab63} that the pairing 
singularity is logarithmic and unimportant for integrated quantities.  
A more quantitative statement is needed to have complete understanding
of low--density fermionic matter. 

Another motivation for our study is the general reexamination of nuclear 
physics with effective field theory which is now taking place 
\cite{Weinberg1,KoMany,Parka,LMa,GPLa,Bvk,KSW,Hsu}. 
In the effective field theory approach, 
the interaction is systematically expanded in a power series in
momentum with the object of getting relationships between observables
such that the details of the short-distance interaction
need not be parameterized.  We shall show here that the BCS theory of
pairing is amenable to this approach, and the low--energy theory
gives finite and analytic results. 
Within effective field theory many results
can be obtained analytically opposed to the numerical treatment of potential 
models. 
In this sense our approach complements the large body of literature
of pairing in nuclear and neutron matter that is based on potential 
models \cite{Wambach89,Baldo90,Chen93,Takat93,Khodel96,Carlson97,Elgaroy}.

We consider a Fermi gas with two-fold degeneracy interacting with
a short-range attractive interaction.  Examples are neutron
matter or gaseous $^3$He.  The Hamiltonian is idealized to be
of the form 
\def\intk{\int {d^3k\over(2\pi)^3}}
\be
\label{ham}
H = V\intk \epsilon_k \left(a_{k,\uparrow}^\dagger a_{k,\uparrow}
             +a_{-k,\downarrow}^\dagger a_{-k,\downarrow}\right)
+ gV^2 \intk {\int {d^3k'\over(2\pi)^3}}a_{k,\uparrow}^\dagger 
a_{-k,\downarrow}^\dagger
a_{-k',\downarrow}a_{k',\uparrow}
\ee
where ${\epsilon_k=k^2/2m}$ is the kinetic energy and $V$ the volume.
In effective field theory the contact interaction is the leading term in a 
derivative expansion of the many--body system. This limits the validity of
the Hamiltonian (\ref{ham}) to the regime of long wave lengths
or small densities. However, corrections can systematically be implemented. 
We have only retained terms in the contact interaction
that are needed in the wave function.  The BCS wave function has
the form $ |\Psi\rangle = \Pi_k (U_k + V_k a_{k,\uparrow}^\dagger
a_{-k,\downarrow}^\dagger)|0\rangle$; the energy is minimized with respect
to $U_k,V_k$ to get the BCS equations \cite{BCS}. The equation for the pairing 
gap $\Delta$ is
\be
\label{gapeq}
1 = -{gV\over 2 (2\pi)^3} \int {d^3k\over\sqrt{(\epsilon_k-\lambda)^2 
+\Delta^2}
},
\ee
where $\lambda$ is the chemical potential. The density is given
in terms of these parameters by
\be
\label{densdef}
{N\over V} =  \int {d^3 k\over (2 \pi)^3} \Bigg[1-{\epsilon_k-\lambda\over
\sqrt{(\epsilon_k-\lambda)^2+\Delta^2}}\Bigg].
\ee
Finally, the energy density of the paired state is given by
\be
\label{edens}
{E \over V} =  \int {d^3 k \over (2\pi)^3} \Bigg[\epsilon_k -
{\epsilon_k(\epsilon_k-\lambda)\over\sqrt{(\epsilon_k-\lambda)^2+\Delta^2}}
-{1\over 2}{\Delta^2\over\sqrt{(\epsilon_k-\lambda)^2+\Delta^2}}
\Bigg].
\ee
Note that the last two integrals are finite, although each integrand is a 
sum of terms that are individually divergent.  

The problem with Eq.~(\ref{gapeq}) as derived is that the contact interaction 
is singular in three dimensions.  One often introduces a cutoff to 
make the integrals converge. However, in effective field theory cutoffs 
are not explicitly introduced.  Rather, the computed observables are 
expressed directly in
terms of other physical quantities.  To leading order in a low--energy
expansion of the interaction, the physical quantity is the scattering
length.  With the same Hamiltonian, the scattering 
length $a$ is given by a similar divergent integral,
\be
\label{scattlen}
-{mgV\over4\pi a}+1 = -{gV\over 2 } \int {d^3k\over (2\pi)^3}{1\over\epsilon_k}
\ee
Let us now subtract equations (\ref{gapeq}) and (\ref{scattlen}) to obtain
\be
\label{master}
{mg\over 4\pi a} = - {g\over 2 (2\pi)^3} \int d^3k\left[{1\over
\sqrt{(\epsilon_k-\lambda)^2 
+\Delta^2}}-{1\over \epsilon_k}\right].
\ee
Notice that the integral is now convergent and so any cutoff can be taken
to infinity. Furthermore, the strength of the contact interaction $g$, 
which is also an unphysical quantity, can be divided out.    
It is convenient to evaluate both terms of the integral (\ref{master}) 
separately by dimensional regularization (DR) \cite{Collins}. In
DR, integrals of powers are zero so the second term in the integrand in
(\ref{master}) can be dropped. The first term can be evaluated using 
\cite{Gradshteyn} (3.252.11),
\be
\label{integral}
\int\limits_0^\infty dz\,\frac{z^\alpha}{\left[(z-1)^2+x^2\right]^{1/2}}
 = -{\pi\over \sin\pi\alpha}\,(1+x^2)^{\alpha/2}\,
P_\alpha\left(-1/\sqrt{1+x^2}\right),
\ee
where $P_\alpha$ denotes the Legendre function.

\begin{figure}
  \begin{center}
    \leavevmode
    \parbox{0.9\textwidth}
           {\psfig{file=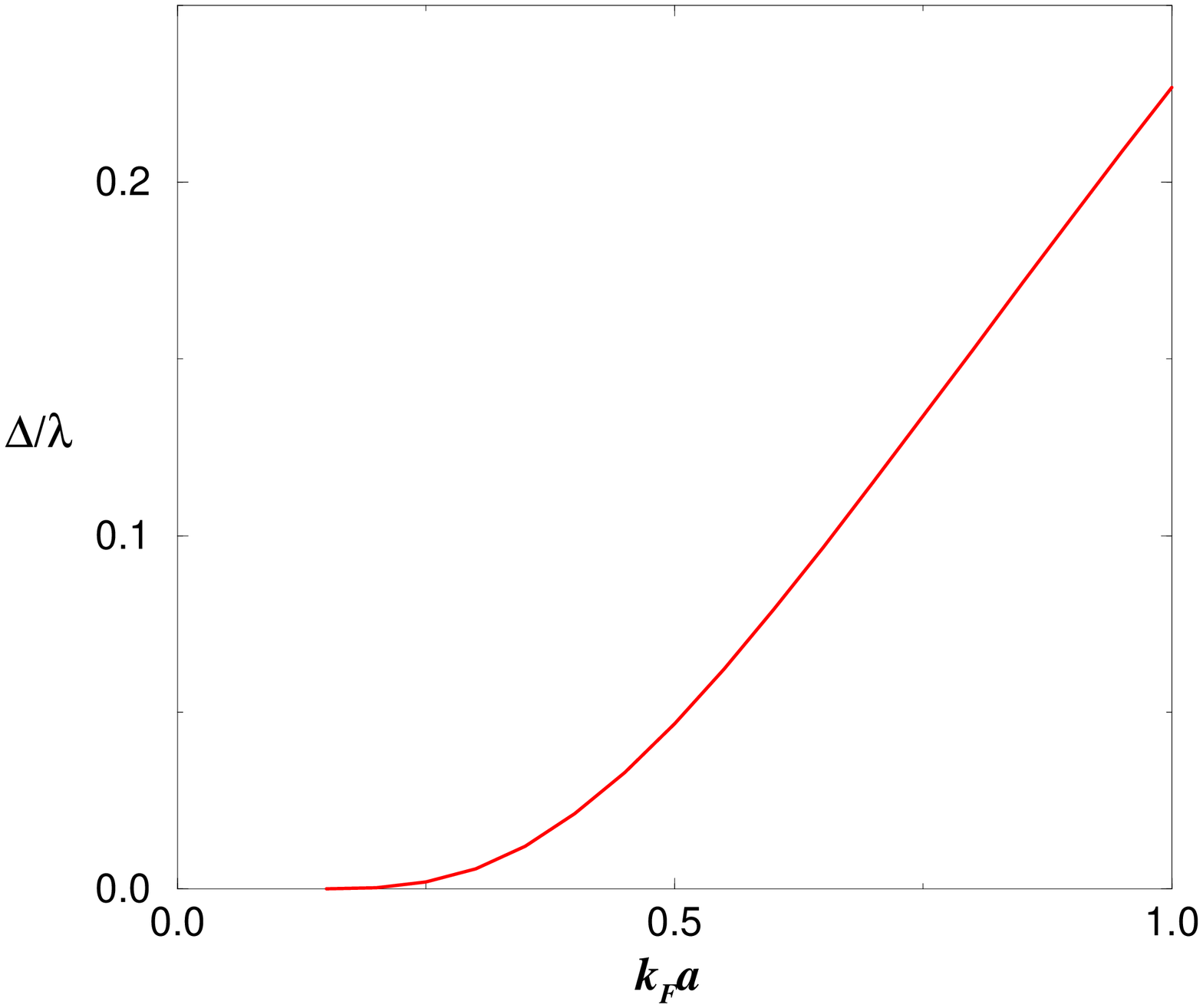,width=0.9\textwidth,angle=270}}
  \end{center}
\protect\caption{Energy gap $\Delta/\lambda$ as a function of $k_F a$}
\label{fig1}
\end{figure}

We write the final result in the form
\be
\label{scattgap}
{1\over k_F a} = (1+x^2)^{1/4} \,P_{1/2}\left(-1/\sqrt{1+x^2}\right)
\ee
where $k_F = \sqrt{2 m\lambda}$ is the Fermi momentum and
$x=\Delta/\lambda$.  This is our main result.  Eq.~(\ref{scattgap}) is 
graphed in Fig.~\ref{fig1}. For small values of $k_Fa$ the gap is 
exponentially small as in the usual
BCS theory,
\be
\label{expgap}
\Delta = {8\over e^2}\lambda\exp{\left(-\pi\over 2 k_F |a|\right)}.
\ee 
This comes about by the behavior of $P_{1/2}(z)$, 
which has a logarithmic singularity at $z=-1$ \cite{Bateman}. 
Eq.~(\ref{expgap}) agrees with the result derived in ref. \cite{Khodel96}.
For large values of $k_Fa$, the gap is proportional to $\lambda$, approaching
$\Delta\approx 1.16\lambda$. 

For neutron matter the solution of eq. (\ref{scattgap}) agrees
with numerical results from potential models only for small values 
of the Fermi momentum. The large value of the scattering length ($a=-18.8$fm)
clearly limits the domain of validity of the Hamiltonian (\ref{ham}). 
In the appendix we consider pairing in the effective range approximation.
This improves on the precision of the calculation in the low--density regime
but does not enlarge the domain of validity. 

\begin{figure}
  \begin{center}
    \leavevmode
    \parbox{0.9\textwidth}
           {\psfig{file=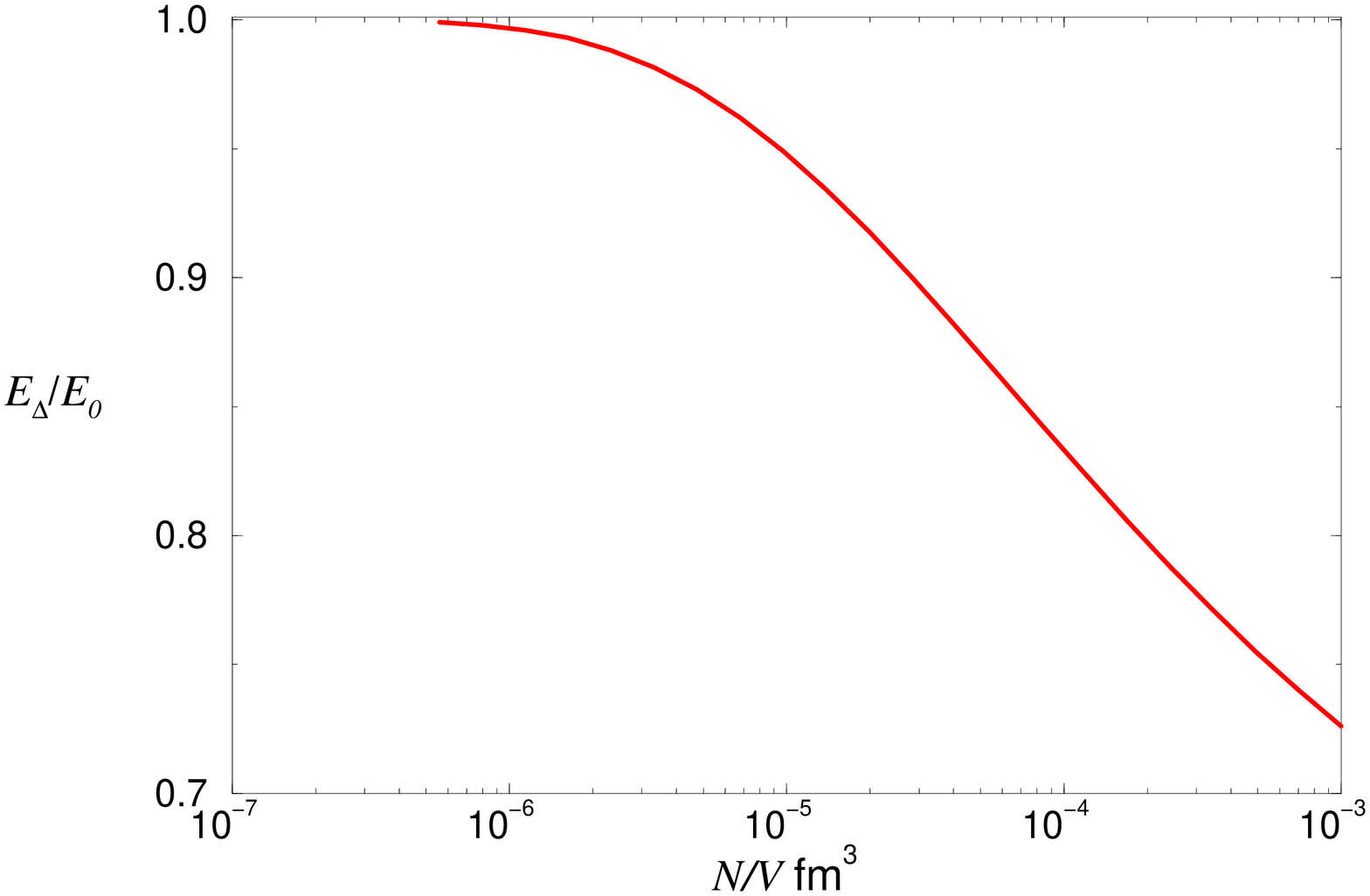,width=0.9\textwidth,angle=270}}
  \end{center}
\protect\caption{Energy $E_\Delta$ of interacting neutron matter normalized to
the energy $E_0$ of noninteracting neutrons as a function of the density.}
\label{fig2}
\end{figure}

We complete this discussion by computing the energy density (\ref{edens}) and
the density (\ref{densdef}). The finite integrals involved are very similar 
to the previous one and can be evaluated using the same DR integral, 
eq.~(\ref{integral}). The density  of the BCS state is given by
\be
\label{dens}
{N\over V} = -{k_F^3 \over 4\pi}\, (1+x^2)^{1/4}
        \left[ P_{1/2}\left(-1/\sqrt{1+x^2}\right)
   + \sqrt{1+x^2} \,P_{3/2}\left(-1/\sqrt{1+x^2}\right)\right],  
\ee

and the energy density by
\be
\label{erg}
{E\over V} = -{3\over 20\pi}k_F^3\lambda (1+x^2)^{1/4}\left[
(1+x^2/6)\,P_{1/2}\left(-1/\sqrt{1+x^2}\right)
+\sqrt{1+x^2}\,P_{3/2}\left(-1/\sqrt{1+x^2}\right)\right]. 
\ee
For fixed density and scattering length eqs.~(\ref{scattgap}) and (\ref{dens})
can be solved for the pairing gap and the Fermi energy. Put into 
eq.~(\ref{erg}) this yields the energy of the interacting system at fixed 
density. Fig.~\ref{fig2} shows a comparison with the energy of 
noninteracting neutrons. For $|k_Fa|\approx 1$ 
(i.e. $N/V\approx 5\times 10^{-6}$ fm$^{-3}$) 
pairing lowers the energy about 3\% confirming the qualitative statement
that the effects of pairing on the binding are mild.

{\em Discussion} --- 
We now discuss the domain of validity of this low--density expansion. 
As pointed out above, the applicability of Hamiltonian (\ref{ham}) is limited 
to the regime of long wave length $|k_Fa|\ll 1$ or small densities. The 
description of neutron or nuclear matter at nuclear densities requires 
the inclusion of the effective range and pions. Comparing with more 
microscopic calculations involving phenomenological potentials it appears that
deviations from the low--density behavior are set by the scattering length.
Similar considerations can be made for $^3$He where the  
scattering length of the Aziz potential \cite{aziz79} is 
large on an atomic scale. Many-body correlation effects will become 
important when $k_F a \approx 1$.  It might be possible to treat them by 
modifying the strength of the pairing and the density of states in 
eq.~(\ref{ham}).
The sign would be to increase the pairing, but we have not attempted to 
calculate these effects.

Another consideration is whether the low--density phase exists for fermionic
systems with attractive scattering lengths.  In the case of $^3$He, a
low--density phase could only be metastable at zero temperature, because
there is a finite binding of the liquid phase.  However, the metastability
could be quite significant, because the minimum size for a bound drop
is thought to be of the order of fifty particles.  Another indication of
the metastability of a low--density phase is the sound velocity in the
scattering length expansion.  Taking the first three terms, the 
sound velocity is positive at all densities, and thus small deviations
from uniformity are energetically unfavorable.
In the case of neutron matter, it is thought that pressure is always
positive as a function of density, so the low--density state would be
stable.

In summary, we have considered the pairing in low--density Fermi systems
within effective field theory. This model independent approach yields
analytical expressions which relate the pairing gap, 
the density and the ground state energy to the scattering length. 
The analytical derivation of these results is quite interesting.

\section*{Appendix}
To include the effective range we add the effective range potential
\be
\label{efr}
g_2V^2\intk {\int {d^3k'\over(2\pi)^3}}(k-k')^2a_{k,\uparrow}^\dagger 
a_{-k,\downarrow}^\dagger
a_{-k',\downarrow}a_{k',\uparrow}
\ee
to the Hamiltonian (\ref{ham}). The gap equation then becomes
\be
\label{rnggap}
\Delta_p = -{V\over 2}\intk\frac{g+g_2(p-k)^2} 
            {\sqrt{(\epsilon_k-\lambda)^2 +\Delta^2_k}} \Delta_k
\ee
and is explicitly momentum dependent. We make the quadratic ansatz 
$\Delta_p=\Delta + p^2\delta$ for the momentum dependence and obtain 
two coupled equations that express $\Delta$ and $\delta$ in terms of 
(divergent) integrals. To deal with the divergencies we observe that
the integrals' dependence on the Fermi momentum is given by 
\be
\intk {k^{2n}\over\sqrt{(\epsilon_k-\lambda)^2 +(\Delta+k^2\delta)^2}}
={k_F^{2n+1}m\over 2\pi^2\lambda}\, J_{n+1/2}(x,y), 
\ee
where $x=\Delta/\lambda$, $y=\delta k_F^2/\lambda$ and $J_{n+1/2}(x,y)$
is the dimensionless function
\be
J_\alpha(x,y)\equiv\int\limits_0^\infty dt\,
{t^\alpha\over\sqrt{(t-1)^2+(x+yt)^2}}.
\ee
In effective field theory an expansion 
in momenta is quite useful \cite{KoMany,KSW}. In what follows we truncate each of the
gap equations to its leading order in the Fermi momentum and obtain
\ba
\label{coupl}
1&=&-{Vgmk_F\over 4\pi^2}\,J_{1/2}(x,y), \nonumber\\
\delta&=&-{Vg_2\Delta mk_F\over 4\pi^2}\,J_{1/2}(x,y).
\ea
Obviously we have $\delta/\Delta=g_2/g$. To make contact with low energy
scattering data we expand the scattering amplitude 
\be
\label{Aham}
{\cal A}(p)=Vg\left[1+Vg\,I(p)+{g_2\over g}p^2 
+\left(Vg\,I(p)\right)^2\right]
\ee
up to quadratic order in momenta. The loop integral is 
\be
I(p)\equiv{1\over 2}\intk {1\over \epsilon_p -\epsilon_k +i\eta}
={m p\over 4\pi^2}\int\limits_0^\infty dt {t^{1/2}\over 1-t+i\eta}
\ee
At low energies the scattering 
amplitude is given in terms of the scattering length $a$ and the effective
range $r_0$
\be
\label{Adata}
{\cal A}(p)={4\pi a\over m}\left[1-iap + (ar_0/2-a^2)p^2\right].
\ee
Note that the divergence of the integral $J_{1/2}(x,y)$ appearing in the 
gap equations (\ref{coupl}) is similar to that of the loop integral $I(p)$
appearing in the expression (\ref{Aham}) for the scattering amplitude. 
Thus, both divergencies may be taken
care off by a a renormalization of the coupling constants $g$ and $g_2$.
We use dimensional regularization to compute the divergent integrals. One 
obtains $I(p)=-i(m/4\pi)p$ and a comparison of (\ref{Aham}) and (\ref{Adata}) 
yields $g_2/g=ar_0/2$ and $g=4\pi a/m$. Finally we have
\be
J_\alpha(x,y)= -{\pi\over\sin\pi\alpha}(1+y^2)^{-1/2}
\left({1+x^2\over 1+y^2}\right)^{\alpha/2}\,P_\alpha(-z),
\ee
where $z= (1-xy)/\sqrt{(1+x^2)(1+y^2)}$. This yields the final results
\ba
\label{gapfin}
{1\over k_Fa}&=&(1+y^2)^{-1/2}\left({1+x^2\over 1+y^2}\right)^{1/4}\,P_{1/2}(-z),
\nonumber\\
y &=& {ar_0\over 2}k_F^2 x.
\ea
Note that these equations add corrections of the order $\sim k_F^2ar_0$ 
to the gap equation ~(\ref{scattgap}). These corrections are small only in the
low--density regime $k_Fa \ll 1$. For a description of neutron matter 
($a=-18.8$fm $r_0=2.75$fm) at larger densities, at least the inclusion of 
pions seems to be necessary. Note also, that 
the gap equations (\ref{gapfin}) become singular for $k_F^2\to -2/ar_0$ 
(i.e. $y\to -x$)
due to the logarithmic singularity of the Legendre function for $z\to -1$.
This behavior results from the quadratic approximation for the 
interaction potential and the truncations in the gap equation. It is related
to the change in sign of the truncated potential at $k_F=\sqrt{-2/ar_0}$ 
\cite{Khodel96}. Again, the introduction of pions or higher potential terms 
seem to be necessary to alter this behavior.

\section*{Acknowledgments}

We acknowledge discussions with P. Bedaque, A. Bulgac, H. Grie{\ss}hammer, 
D. Kaplan and M. Savage. We thank J. Hormuzdiar and S.D.H. Hsu for pointing out
a correction to formula (\ref{expgap}). This work was supported by the Dept. of
Energy under Grant DE-FG-06-90ER-40561.

\begin{references}
\bibitem{ab63} 
A.A. Abrikosov, L.P. Gorkov, and I.E. Dzyaloshinski,
Methods of Quantum Field Theory in Statistical Physics, (Prentice-Hall,1963).
%
\bibitem{Weinberg1}  
S. Weinberg, \pl B {\bf 251}(1990)288; 
Nucl. Phys. B {\bf 363}(1991) 3; \pl B {\bf 295} (1992) 114
%
\bibitem{KoMany}  
C. Ordonez, L. Ray, and U. van Kolck, 
\prl {\bf 72} (1994) 1982; 
%
\bibitem{Parka}  
T.S. Park, D.P. Min and M. Rho, 
\prl {\bf 74} (1995) 4153; Nucl. Phys. A {\bf 596} (1996) 515
%
\bibitem{LMa}  
M. Luke and A.V. Manohar, \prd {\bf 55} (1997) 4129
%
\bibitem{GPLa}  
G.P. Lepage, {\tt nucl-th/9706029}, Lectures given at 9th
Jorge Andre Swieca Summer School: Particles and Fields, Sao Paulo, Brazil,
16-28 Feb 1997.
%
\bibitem{Bvk}  
P.F. Bedaque and U. van Kolck,
\pl B {\bf 428} (1998) 221; 
P.F. Bedaque, H.-W. Hammer and U. van Kolck, 
\prc {\bf 58}  (1998) R641
%
\bibitem{KSW}  
D.B. Kaplan, M.J. Savage and M.B. Wise, 
\pl B {\bf 424} (1998) 390;
{\tt nucl-th/9802075}, {\it to appear in Nucl. Phys. B};
%
\bibitem{Hsu}
S.D.H. Hsu and J. Hormuzdiar,
{\tt nucl-th/9811017}

\bibitem{Wambach89}
T.L. Ainsworth, J. Wambach, and D. Pines,
\pl B {\bf 222} (1989) 173
%
\bibitem{Baldo90} 
M. Baldo, J. Cugnon, A. Lejeune, and U. Lombardo, 
Nucl. Phys. A {\bf 515} (1990) 409
%
\bibitem{Chen93}
J.M.C. Chen, J.W. Clark, R.D. Dav{\'e}, and V.V. Khodel,
Nucl. Phys. A {\bf 555} (1993) 59
%
\bibitem{Takat93}
T. Takatsuka and R. Tamagaki, 
Prog. Theor. Phys. Suppl. {\bf 112} (1993) 27
%
\bibitem{Khodel96}
V.A. Khodel, V.V. Khodel, and J.W. Clark, 
Nucl. Phys. A {\bf 598} (1996) 390
%
\bibitem{Carlson97}
B.V. Carson, T. Frederico, and F.B. Guimaraes,
\prc {\bf 56} (1997) 3097
%
\bibitem{Elgaroy}
\O. Elgar{\o}y and M. Hjorth-Jensen,
\prc {\bf 57} (1998) 1174
%
\bibitem{BCS}
J. Bardeen, L.N. Cooper, and J.R. Schrieffer,
Phys. Rev. {\bf 108} (1957) 1175
%
\bibitem{Collins}
J.C. Collins,
{\it Renormalization},
Cambridge Univ. Press, Cambridge (1984)
%
\bibitem{Gradshteyn}
I.S. Gradshteyn and I.M. Rezhik,
{\it Table of Integrals, Series and Products}
(1980)
%
\bibitem{Bateman}
A. Erd{\'e}lyi (Ed.),
{\it Higher Transcendental Functions}, Vol. I, McGraw Hill, N.Y. (1953) 
%
\bibitem{aziz79} R.A. Aziz, V.P.S. Nain, J.S. Carley, W.L. Taylor, 
and G.T. McConville; 
\jcp {\bf 70} (1979) 4330;
A.R. Janzen and R.A. Aziz,
\jcp {\bf 103} (1995)  9626
%
\end{references}
\end{document}